\documentclass[aps,pra,superscriptaddress,reprint,floatfix,longbibliography,showpacs,amsfonts,amsmath,amssymb]{revtex4-1}
\usepackage{graphicx}
\usepackage{hyperref}
\usepackage{xr-hyper}
\usepackage{upgreek}
\usepackage{epstopdf}
\usepackage[note-name=, use-sort-key = false]{notes2bib}
\usepackage{color}
\usepackage{soul}
\usepackage{ulem}

    \hypersetup{
    	colorlinks = true,
		citecolor = blue,
		bookmarksnumbered = true,
		urlcolor = blue
	}

\begin{document}

\title{Optical anisotropy of the kagome magnet FeSn:
Dominant role of excitations between kagome and Sn layers}

\author{J. Ebad-Allah}
\email{These authors contributed equally to this work.}
\affiliation{Experimentalphysik II, Institute for Physics, Augsburg University, D-86135 Augsburg, Germany}
\affiliation{Department of Physics, Tanta University, 31527 Tanta, Egypt}
\author{M.-C. Jiang}
\email{These authors contributed equally to this work.}
\affiliation{Department of Physics and Center for Theoretical Physics, National Taiwan University, Taipei 10617, Taiwan}
\affiliation{RIKEN Center for Emergent Matter Science, 2-1 Hirosawa, Wako 351-0198, Japan}
\author{R. Borkenhagen}
\affiliation{Experimentalphysik II, Institute for Physics, Augsburg University, D-86135 Augsburg, Germany}
\author{F. Meggle}
\affiliation{Experimentalphysik II, Institute for Physics, Augsburg University, D-86135 Augsburg, Germany}
\author{L. Prodan}
\affiliation{Experimentalphysik V, Center for Electronic Correlations and Magnetism, Institute for Physics,
Augsburg University, D-86135 Augsburg, Germany}
\affiliation{Institute of Applied Physics, Moldova State University, MD-2028 Chisinau, Republic of Moldova}
\author{V. Tsurkan}
\affiliation{Experimentalphysik V, Center for Electronic Correlations and Magnetism, Institute for Physics,
Augsburg University, D-86135 Augsburg, Germany}
\affiliation{Institute of Applied Physics, Moldova State University, MD-2028 Chisinau, Republic of Moldova}
\author{F. Schilberth}
\affiliation{Experimentalphysik V, Center for Electronic Correlations and Magnetism, Institute for Physics, Augsburg University, D-86135 Augsburg, Germany}
\affiliation{Department of Physics, Institute of Physics, Budapest University of Technology and Economics, M\H{u}egyetem rkp. 3., H-1111 Budapest, Hungary}
\author{G.-Y. Guo}
\affiliation{Department of Physics and Center for Theoretical Physics, National Taiwan University, Taipei 10617, Taiwan}
\affiliation{Physics Division, National Center for Theoretical Sciences, Taipei 10617, Taiwan}
\author{R. Arita}
\affiliation{RIKEN Center for Emergent Matter Science, 2-1 Hirosawa, Wako 351-0198, Japan}
\affiliation{Research Center for Advanced Science and Technology, University of Tokyo, 4-6-1 Meguro-ku, Tokyo 153-8904, Japan}
\author{I. K\'{e}zsm\'{a}rki}
\affiliation{Experimentalphysik V, Center for Electronic Correlations and Magnetism, Institute for Physics,
Augsburg University, D-86135 Augsburg, Germany}
\author{C. A. Kuntscher}
\affiliation{Experimentalphysik II, Institute for Physics, Augsburg University, D-86135 Augsburg, Germany}

\begin{abstract}
Antiferromagnetic FeSn is considered to be a close realization of the ideal two-dimensional (2D) kagome lattice, hosting Dirac cones, van Hove singularities, and flat bands, as it comprises Fe$_3$Sn kagome layers well separated by Sn buffer layers.
We observe a pronounced optical anisotropy, with the low-energy optical conductivity being surprisingly higher perpendicular to the kagome planes than along the layers. This finding contradicts the prevalent picture of dominantly 2D electronic structure for FeSn.
Our material-specific theory reproduces the measured conductivity spectra remarkarbly well. A site-specific decomposition of the optical response to individual excitation channels shows that the optical conductivity for polarizations both parallel and perpendicular to the kagome plane is dominated by interlayer transitions between kagome layers and adjacent Sn-based layers. Moreover, the matrix elements corresponding to these transitions are highly anisotropic, leading to larger out-of-plane conductivity.  Our results evidence the crucial role of interstitial layers in charge dynamics even in seemingly 2D systems.
\end{abstract}

\pacs{}

\maketitle
The kagome lattice is a two-dimensional (2D) hexagonal network of corner-sharing triangles and is heavily investigated recently due to the variety of emergent quantum phases such as quantum spin liquid, Dirac or magnetic Weyl fermions, and magnetic skyrmions \cite{Yan.2011, Broholm.2020, Mazin.2014, Kuroda.2017, Hou.2017}. In its tight-binding band structure, the kagome lattice hosts a saddle point, leading to van Hove singularities in the density of states, two linearly dispersing Dirac-bands around the $K$ point, that can give rise to topologically nontrivial Weyl points and nodal lines, and a flat band, emerging from destructive interference of electronic wavefunctions. Due to this interference, the electronic states become geometrically confined within a kagome hexagon \cite{Xu.2020, Yin.2019, Kang.2019, Ye.2018, Armitage.2018, Bergholtz2013, Kang2020, Meier.2020}. This localization and the related quench of kinetic energy amplify electronic correlations, making the kagome lattice an ideal playground to study the interplay between geometry, topology, and correlation physics.

One fundamental question is how these prototypical states evolve upon the stacking of kagome layers in three-dimensional (3D) crystals. In other words, whether the characteristic features of the kagome band structure are still manifested in real materials. For example, it was shown for the bilayer kagome material Fe$_3$Sn$_2$ that the interplane hopping causes a double Dirac structure near $K$, with Dirac points extending to helical nodal lines, and the flat bands acquire dispersion \cite{Ye.2018,Fang.2022, Schilberth.2022}. As was recently shown for the sister compound Co$_3$Sn$_2$S$_2$, such gapped nodal lines can leave fingerprints in the optical response and dominate the anomalous Hall conductivity \cite{Okamura2020, Schilberth2023}. Also, in 3D flat band materials like the pyrochlore metal CaNi$_2$, containing stacked kagome layers, deviations from flatness occur due to multiple vectorial interferences of several Ni $3d$ orbitals \cite{Wakefield.2023}.
Therefore, sensitive spectroscopic probes are required both for the identification of these states and for determining their impact on the materials response, e.~g.,~transport or optical properties.

Our target material is the kagome metal FeSn with antiferromagnetic (AFM) order below $T_\text{N}$=365~K \cite{Yamamoto1966, Sales2019}. It has been suggested to host Dirac fermions and flat bands, exhibiting characteristic features of an individual kagome lattice despite the interlayer interactions inevitably present in 3D crystals.  \cite{Kang.2019, Multer.2023, Han2021}. It crystallizes in the hexagonal space group P6/$mmm$, where Fe$_3$Sn kagome layers in the $ab$ plane are alternatingly stacked with Sn honeycomb layers along the $c$ direction \cite{Kang.2019}. FeSn forms a kagome metal series together with Fe$_3$Sn$_2$ and Fe$_3$Sn. In this order, the electronic structure is claimed to turn from dominantly 2D to 3D, with FeSn being the closest realization of the independent kagome layers \cite{Kang.2019}.

Electronic band structure calculations of FeSn indeed predicted flat bands around 0.5~eV above the Fermi energy $E_\text{F}$ and a Dirac cone located at the $K$ point around 0.4~eV below $E_\text{F}$, together with a corresponding nodal line along $K-H$ \cite{Lin.2020, Fang.2022, Zhang.2022,Kang.2019}. Furthermore, flat bands in limited regions of the Brillouin zone were predicted \cite{Lin.2018,Fang.2022}, as well as observed by angle resolved photoemission spectroscopy (ARPES) and scanning tunneling spectroscopy, in both FeSn and Fe$_3$Sn$_2$ \cite{Kang.2019, Multer.2023}. ARPES measurements also revealed the predicted Dirac cone(s) in FeSn and Fe$_3$Sn$_2$, respectively \cite{Kang.2019,Ye.2018}. For Fe$_3$Sn$_2$ it was shown that electronic correlation effects cannot be neglected, since they cause a reorganization of the energy bands close to $E_\text{F}$ \cite{Schilberth.2022}. It should also be noted that there are several additional conventional energy bands in the vicinity of $E_\text{F}$ according to electronic band structure calculations \cite{Schilberth.2022}. Nevertheless, recent optical spectroscopy studies on Fe$_3$Sn$_2$ observed features in the low-energy optical conductivity spectrum, which were interpreted as excitations of Dirac cones, helical nodal lines and flat bands close to $E_\text{F}$ \cite{Biswas.2020, Schilberth.2022}.

In this Letter, we report the anisotropic optical conductivity of the kagome magnet FeSn based on a combined experimental and theoretical study. We find a pronounced anisotropy in the low-energy optical conductivity, with a higher magnitude perpendicular to the kagome plane. This is opposite to the general expectations,  where the reduced dimensionality of the crystal structure is directly manifested in the optical anisotropy \cite{Kezsmarki2006}. By comparing the experimental optical conductivity with material-specific theory calculated in the framework of density functional theory we can attribute the spectral features to specific intra- and interlayer excitations, revealing significant contributions of the interstitial Sn layers to both in- and out-of-plane conductivity.  No clear sign of flat band transitions and Dirac electron excitations is detected in the optical conductivity of FeSn.

\begin{figure}[t]
\includegraphics[width=0.43\textwidth]{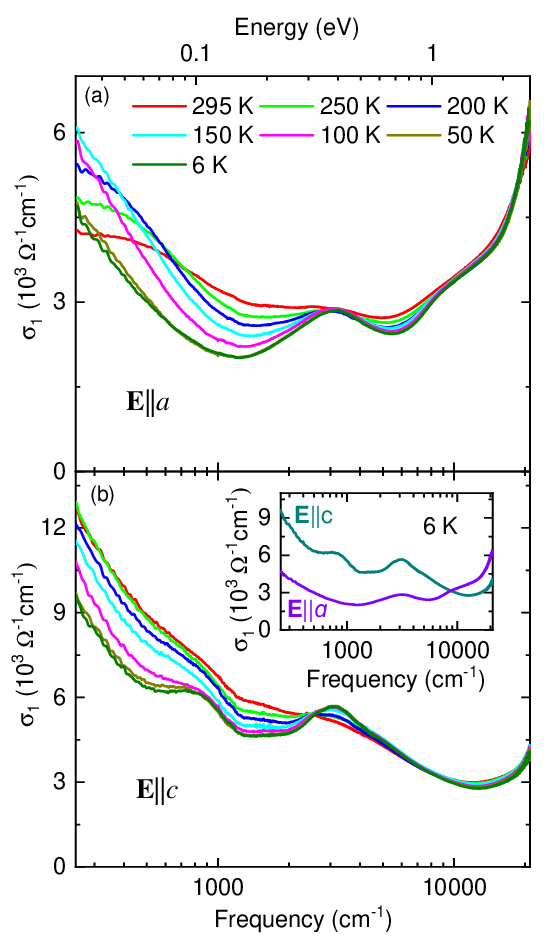}
\caption{Experimental optical conductivity spectra $\sigma_{1}$ of FeSn for polarization direction (a) {\bf E}$\parallel$$a$ and (b) {\bf E}$\parallel$$c$ as a function of temperature. Inset of (b): Comparison of {\bf E}$\parallel$$a$ and {\bf E}$\parallel$$c$ optical conductivity at 6~K.}
\label{fig:conductivity}
\end{figure}

A comparison of the experimental optical conductivity spectra of single-crystalline FeSn for the polarization directions {\bf E}$\parallel$$a$ (parallel to the kagome plane) and {\bf E}$\parallel$$c$ (perpendicular to the kagome plane) is presented in Fig.\ \ref{fig:conductivity} (a) and (b), respectively.
The experimental details about the crystal growth, the reflectivity measurements and the corresponding polarization- and temperature-dependent reflectivity spectra can be found in the supplemental material (SM)~\cite{Suppl} (see also references \cite{EbadAllah.2019,Koepf.2020,EbadAllah.2021,Tanner.2015,Kresse1996a,
Kresse1996b,Kresse1999,Perdew1996,Pizzi2020,Schilberth.2022,Biswas.2020,
Li2023} therein).
For {\bf E}$\parallel$$a$ we observe a pronounced absorption band at around 3100~cm$^{-1}$ ($\sim$0.38~eV) and a dip in the frequency range 1000 - 1300~cm$^{-1}$ ($\sim$0.12 - 0.16~eV) followed by the low-frequency Drude contribution. The metallic Drude response is also resolved for {\bf E}$\parallel$$c$, in fact, it is stronger than in the in-plane conductivity. The out-of-plane conductivity spectrum also shows an absorption band around 3100~cm$^{-1}$ and an additional one around 800~cm$^{-1}$. For both polarization directions the observed absorption bands hardly shift with decreasing temperature but only sharpen.
Most interestingly, for energies below 9000 cm$^{-1}$ ($\sim$1.1~eV) the optical conductivity perpendicular to the kagome layers is {\it higher} than along the kagome plane, as illustrated in the inset of Fig.\ \ref{fig:conductivity}(b).
These results clearly demonstrate a strong anisotropy of charge excitations in FeSn, but in the opposite way as one would expect for independent kagome layers.

This behavior contradicts the frequently made assumption that transport in layered systems is mostly confined to the individual layers. In fact, similar behavior has been recently reported  for the
dc transport characteristics in the
sister compound CoSn \cite{Huang2022}. Huang \textit{et al.} attributed the observed anomalous dc transport in CoSn to the unique properties of flat band electrons with large in-plane effective mass. In addition, a larger dc conductivity along the $c$ direction was also reported for other layered kagome magnets such as HoAgGe \cite{Zhao2020}, YCr$_6$Ge$_6$ \cite{Yang2022}, Fe$_3$Sn \cite{Prodan.2023}, Fe$_3$Sn$_2$ \cite{Du2022}, and recently suggested for  FeSn \cite{Sales2019}. The last three examples show that this ``reversed'' conductivity anisotropy is a general attribute of iron-tin kagome magnets.

\begin{figure*}[t]
\includegraphics[width=1\textwidth]{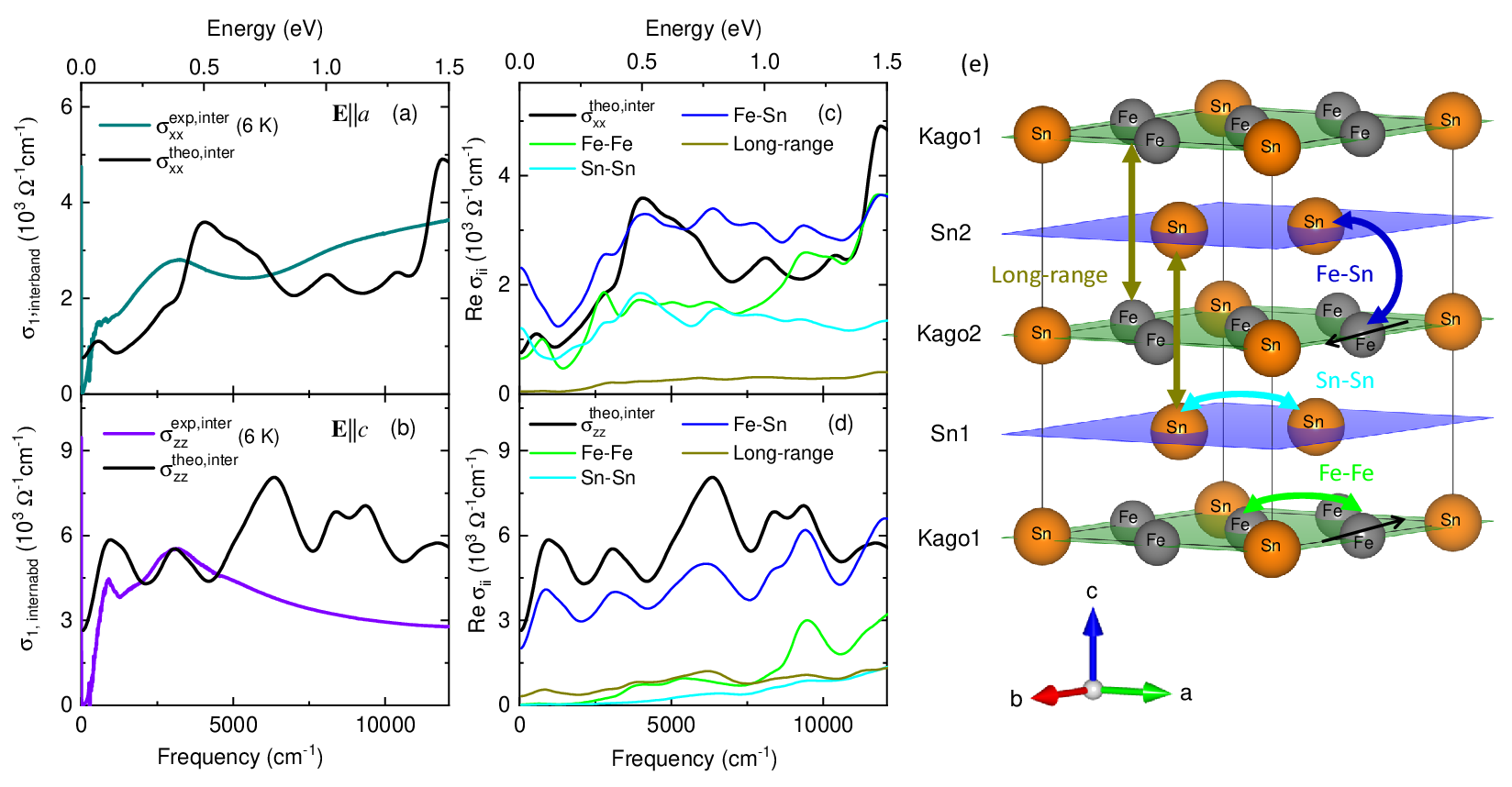}
\caption{Comparison between the experimental and theoretical interband optical conductivity spectra of FeSn for polarization direction (a) {\bf E}$\parallel$$a$ and (b) {\bf E}$\parallel$$c$. Panels (c) and (d) show the site-specific contributions to the calculated $\sigma_{xx}$ and $\sigma_{zz}$, namely intralayer excitations within the Fe kagome plane (Fe-Fe), intralayer excitations within the Sn plane (Sn-Sn), and interlayer excitations between the Fe kagome plane and the Sn plane (Fe-Sn). The intralayer and interlayer excitations are illustrated in (e) showing the crystal structure of FeSn. The
Fe$_3$Sn kagome layers and the Sn buffer layers are highlighted in green and blue, respectively. The two black arrows indicate the A-type antiferromagnetic order of Fe magnetic moments, as described in the text.}
\label{fig:comparison-exptheory}
\end{figure*}

For the separation of intra- and interband transitions in the optical conductivity spectra, we carried out simultaneous fittings of reflectivity and optical conductivity data using the Drude-Lorentz model. Before investigating the interband excitations, utilizing the results of the intraband excitations, we will briefly comment on the estimation of the electronic correlation strength, which is often discussed for kagome materials~\cite{Schilberth.2022}. From optical conductivity, we can estimate the electronic correlation strength by calculating the ratio between the electronic kinetic energy derived from the experimentally measured Drude weight and the calculated plasma frequency from the band theory ($\omega_{\text{pl},ii}$)~\cite{Dagotto1994,Millis2005,Qazilbash2009}. The plasma frequency is directly linked to the Drude spectral weight as described in Ref.\ \cite{Dressel.2002}.
The results of the $\omega_{\text{pl},ii}$ from the experiments (calculations) are 1.62 eV (2.40 eV) and 2.38 eV (2.06 eV) for $\omega_{\text{pl},xx}$ and $\omega_{\text{pl},zz}$, respectively. Details are given in the Supplemental Material  (SM)~\cite{Suppl}. We can see that in both the in-plane and out-of-plane direction, the plasma frequency is in the same order of magnitude, making the electron kinetic energy ratio expected to fall around 1. This means that FeSn is a conventional metal, i.e., it is weakly correlated.

Here, we refocus back on the contributions of interband excitations to the optical conductivity and therefore subtract the Drude contributions from the measured conductivity spectra. The resulting interband optical conductivity $\sigma_{1,\text{interband}}$ at 6~K is depicted in Figs.\ \ref{fig:comparison-exptheory}(a) and \ref{fig:comparison-exptheory}(b) for {\bf E}$\parallel$$a$ and {\bf E}$\parallel$$c$, respectively (the fit of the $\sigma_1$-spectra with all fitting contributions can be found in the SM~\cite{Suppl}). For both polarization directions, we find two distinct absorption bands in the energy range below $\sim$0.6~eV, centered at around 80~meV (105~meV) and 380~meV (380~meV) for  {\bf E}$\parallel$$a$ ({\bf E}$\parallel$$c$), respectively. The higher-energy transitions cause a broad excitation continuum in the conductivity profile
for both polarizations.

To understand the pronounced anisotropy in the optical response of FeSn and to specify the excitations contributing to its interband optical conductivity (IOC), we carried out \textit{ab initio} calculations in the framework of density functional theory. Note that the calculations are conducted  for an A-type AFM FeSn [i.e., magnetic moments of the Fe atoms are ferromagnetically aligned within each kagome plane but AFM-coupled along the $c$ axis, see Fig.\ \ref{fig:comparison-exptheory}(e)] without considering spin-orbit coupling. Fig.~S4 in the SM~\cite{Suppl} displays the results with and without the spin-orbit coupling, and we can see that the spin-orbit coupling has a minor effect on the diagonal part of IOC in FeSn. The IOC can be obtained from the Kubo-Greenwood formula written as follows \cite{Pizzi2020}:
\begin{multline}
\noindent\sigma_{\textbf{k},\alpha\beta}(\hbar\omega)=\frac{i e^2}{\hbar V}\sum_{\textbf{k},n,m}(f_{m,\textbf{k}}-f_{n,\textbf{k}})\cdot\\
\frac{\epsilon_{m,\textbf{k}}-\epsilon_{n,\textbf{k}}}{\epsilon_{m,\textbf{k}}-\epsilon_{n,\textbf{k}}-(\hbar\omega+i\eta)}A^{\rm (H)}_{nm,\alpha}(\textbf{k})A^{\rm (H)}_{nm,\beta}(\textbf{k})
\label{eq:Kubo}
\end{multline}
where
\begin{equation}
A^{\rm (H)}_{nm,\alpha} = \left\langle u_{n,\textbf{k}} \middle | i \nabla_{\textbf{k}_\alpha} \middle |u_{m,\textbf{k}}\right\rangle
\label{eq:Berryconnec}
\end{equation}
is the Berry connection written in the Hamiltonian gauge.
The $\alpha$ and $\beta$ are the indices in the Cartesian coordinates, $V$ is the cell volume, $f_{n,\textbf{k}} = f(\epsilon_{n,\textbf{k}})$ is the Fermi-Dirac distribution function, $\omega$ is the optical frequency and $\eta>0$ is the smearing parameter. Under the Wannier interpolation, every object inside Eqn.\ \ref{eq:Berryconnec} should consistently be in either the Hamiltonian gauge ($A^{\rm (H)}$) or the Wannier gauge ($A^{\rm (W)}$) \cite{Wang2006}. The Hamiltonian gauge labels $n, m$ as band indices of the projected band structures in the Hilbert space, while the Wannier gauge labels $n, m$ as state vectors in the ``tight-binding space'' defined by the real-space Wannier functions \cite{Wang2006}. A unitary rotation matrix connects these two gauges. Importantly, this allows us to locate particular transitions not only in and/or associate them with by energy bands in $k$-space (after the similarity transformation), but also by the real-space Wannier functions (before the similarity transformation).

First of all, we highlight that the IOCs, including their spectral shape and anisotropy, are nicely captured by the theoretical calculations, as shown in Figs.~\ref{fig:comparison-exptheory}(a), (b). Peaks at 0.05, 0.34 eV for $\sigma_{xx}$ (in-plane polarization) and 0.1, 0.36 eV for $\sigma_{zz}$ (out-of-plane polarization) coincide well with the experimental data, respectively. The largest predicted optical anisotropy $\sigma_{zz}/\sigma_{xx}$ in the low-frequency region is 7.37 at 0.13 eV. The ratio $\sigma_{zz}/\sigma_{xx}$ persists to be greater than one up to 1.67 eV.

In order to assign spectral features and to evaluate the origin of the optical anisotropy, we select excitations by real-space orbital sites, as there is an overwhelming amount of possible transitions in the metallic bands of FeSn (see Fig.~S7 in the SM~\cite{Suppl}). Therefore, we can distinguish intralayer excitations between sites within the Fe-kagome- and within the Sn-planes from interlayer excitations between the Fe-kagome-plane and the adjacent Sn-buffer plane, as illustrated in Fig.~\ref{fig:comparison-exptheory}(e).
Note that the long-range interlayer excitations between the two distinct kagome planes and between the two distinct Sn planes [labeled "Long-range" in Fig.~\ref{fig:comparison-exptheory}(e)] are relatively small.
The results are summarized in Figs.~\ref{fig:comparison-exptheory}(c) and (d), where the black line denotes the overall IOC for $\sigma_{xx}$ and $\sigma_{zz}$ calculated via Eqn.~\ref{eq:Kubo} and the others are the site-selected IOC.
In certain frequency windows, the individual contributions, especially for the Fe-Sn excitation, are larger than the total IOC. We would like to address this issue before looking into the site-selected IOC in more detail.

Instead of a truncated integration by omitting some band indices, what we manipulate during the site selection is the Berry connection $A^{\rm (W)}_{nm,\alpha}$ itself. Since the Berry connection is a complex matrix, after the unitary transformation and the complex conjugation of $A^{\rm (H)}_{nm,\alpha}A^{\rm (H)}_{mn,\beta}$, matrix elements may cancel each other. Therefore, blocking certain terms during the site selection may lead to an optical spectrum larger than the total spectrum. Alternatively, we could picture the optical excitations in real space. Excitation due to a certain frequency photon excites not just one but various sites simultaneously. The final measurement is the result of the interference between all these signals. Blocking out terms before the complex conjugation $A^{\rm (H)}_{nm,\alpha}A^{\rm (H)}_{mn,\beta}$ ignores such interference. Thus, site-dependent IOC can be larger than the total IOC. Such a scenario works better in low-energy regions where possible excitations are limited, as we compare the data of $\sigma_{zz}$ below and above 1.4~eV.

With this clarification, we notice that for $\sigma_{zz}$ the interlayer excitation Fe-Sn dominates the optical spectrum below 1.5 eV, as would be expected for transitions for this polarization.
Surprisingly, this contribution is also the largest for the in-plane conductivity $\sigma_{xx}$: The interlayer excitations are dominant between 0.45 and 1.0~eV and are still larger below 0.45 eV, where the intralayer excitations are of the same order. This indicates that in-plane polarization would excite all possible transitions, even favoring Fe-Sn interlayer transitions in certain energy windows, while $z$-polarized light ({\bf E}$\parallel$$c$) would highly favor the transitions between the Fe-kagome plane and the Sn-buffer plane. While most investigations focus on the kagome plane, this finding clearly reveals the crucial role of interstitial layers in the optical properties of kagome materials.

Next, we discuss whether the characteristic signatures of interband transitions of Dirac cones or excitations between flat bands can be observed in the profile of our optical conductivity spectra.
Interband transitions of linearly dispersing bands close to the Dirac/Weyl nodes are expected to cause a power-law behavior in the optical conductivity according to
$\sigma_1$$\sim$$\omega^{d-2}$, where $d$ is the dimensionality of the system \cite{Tabert.2016}.
Flat band excitations should lead to sharp peaks in the $\sigma_1$ spectrum.
Indeed, the 3D Weyl semimetal Co$_3$Sn$_2$S$_2$ with Co$_3$Sn kagome layers shows both a linear-in-frequency behavior and a sharp low-energy peak in its optical conductivity \cite{Xu.2020, Yang2020, Schilberth2023}.
For the kagome magnet TbMn$_6$Sn$_6$, a flat behavior of $\sigma_1$ due to quasi-2D Dirac bands was observed \cite{Li2023}. For Fe$_3$Sn$_2$, signatures of interband transitions of the double Dirac structure and flat band excitations in the optical conductivity were reported \cite{Biswas.2020} but could not be confirmed \cite{Schilberth.2022}.

For FeSn, the most pronounced contributions to the measured IOC are two absorption peaks below 0.6~eV present for both polarization directions [see Figs.\ \ref{fig:comparison-exptheory}(a) and (b)]. It is important to note that a linear-in-frequency (3D Weyl/Dirac cone) or frequency-independent (2D Weyl/Dirac cone) behavior of $\sigma_1$ is not observed in our data.
According to the electronic band structure of FeSn depicted in Fig.\ S7(a) in the SM~\cite{Suppl}, a Dirac nodal line is located at 0.4~eV below $E_\text{F}$ at the $K-H$ line. Furthermore, flat bands occur in only limited $k$-space regions at -0.2~eV or considerably further away from $E_\text{F}$, consistent with earlier reports \cite{Lin.2020}.
In Fig.~S7(b) \cite{Suppl}, we show the band-selected IOC of various band sets. We highlight the band-selected IOC from band 30 to band 32 to address the excitation between Dirac nodal lines. We notice a flat band-selected IOC for $\sigma_{xx}$ from 0.42 eV to 0.61 eV, similar to the feature reported in \cite{Li2023}, which could be related to the transitions between Dirac nodal lines and linearly dispersed bands around the $H$ point and along the $\Gamma-A$ line. Nevertheless, neither $\sigma_{xx}$ nor $\sigma_{zz}$ from band 30 to 32 contributes significantly to the spectrum. In addition, the results with and without spin-orbit coupling are similar, meaning that the gap opening of Dirac nodal lines has minor effects on the interband optical transition (Fig.~S6). We thus conclude that no dominant optical signature of flat bands and Dirac fermions are found in the optical spectra of FeSn as they are overwhelmed by transitions between trivial bands.

Finally, we remark on the influence of AFM order on the optical conductivity by comparing the electronic structures and IOC between nonmagnetic and AFM FeSn. While transitioning from nonmagnetic to AFM FeSn, band splitting happens due to the magnetic moments, as pointed out in Fig. S8(a)~\cite{Suppl}. Spin channels of the same kagome layer lose spin-degeneracy but form degeneracy with the neighboring kagome layer of the opposite spin to maintain zero net magnetization, shown in Fig. S8(b)~\cite{Suppl}. IOC of nonmagnetic FeSn is given in Fig. S9~\cite{Suppl}. The mentioned band splitting is the main factor of the changes in the IOC as we observe similar spectral patterns but peak shifts for $\sigma_{xx}$ and $\sigma_{zz}$. Nevertheless, despite the peak shifting, the optical anisotropy of $\sigma_{zz} > \sigma_{xx}$ is still observed in the nonmagnetic phase, consistent with the fact of overwhelming trivial bands plus the dominant role of kagome to Sn excitations in our target system FeSn.


In conclusion, our combined experimental and theoretical study of the kagome magnet FeSn reveals a pronounced anisotropy in the optical response, with a surprisingly higher optical conductivity for the polarization perpendicular to the kagome layers than along the layers for energies below $\sim$1.1~eV. According to our material-specific calculations, the main contributions to the low-energy optical conductivity stem from interlayer transitions between the Fe-kagome layers and the adjacent Sn-buffer layers for both polarization directions parallel and perpendicular to the kagome plane. The optical conductivity does not reveal the typical signatures of interband transitions of linearly dispersing bands close to the Dirac/Weyl nodes and flat band excitations.

Our study shows that the nature of charge dynamics in FeSn, as revealed in the real compound, is in strong contrast to what is expected for this material on the kagome-model-system level. In general, one should be cautious when deriving or interpreting material properties based on certain structural motifs, such as the kagome layers, instead of investigating the system in its full complexity.

\begin{acknowledgments}
The authors are grateful to Joachim Deisenhofer for fruitful discussions.
This work was supported by the Deutsche Forschungsgemeinschaft (DFG, German Research Foundation) -- TRR 360 -- 492547816. VT acknowledges the the support via Project No. ANCD 20.80009.5007.19.
M.-C. J. was supported by RIKEN's IPA Program. M.-C. J. and G.-Y. G. acknowledge the support from the Ministry of Science and Technology and the National Center for Theoretical Sciences (NCTS) of the R.O.C. R. A. was supported by a Grant-in-Aid for Scientific Research (No. 19H05825) from the Ministry of Education, Culture, Sports, Science and Technology.
\end{acknowledgments}

\end{document}